\documentclass[aip,jcp,twocolumn,amsfonts,amssymb]{revtex4}

\usepackage{graphicx}
\usepackage{amssymb}
\usepackage{amsmath}

\begin{document}

\title{Wave-function and density functional theory studies of dihydrogen complexes}

\author{E. Fabiano}
\affiliation{National Nanotechnology Laboratory (NNL), Istituto Nanoscienze-CNR,
Via per Arnesano 16, 73100 Lecce, Italy}
\affiliation{Center for Biomolecular Nanotechnologies @UNILE, Istituto Italiano di Tecnologia (IIT), Via Barsanti, 73010 Arnesano (LE), Italy}
\author{L. A. Constantin}
\affiliation{Center for Biomolecular Nanotechnologies @UNILE, Istituto Italiano di Tecnologia (IIT), Via Barsanti, 73010 Arnesano (LE), Italy}
\author{F. Della Sala}
\affiliation{National Nanotechnology Laboratory (NNL), Istituto Nanoscienze-CNR,
Via per Arnesano 16, 73100 Lecce, Italy}
\affiliation{Center for Biomolecular Nanotechnologies @UNILE, Istituto Italiano di Tecnologia (IIT), Via Barsanti, 73010 Arnesano (LE), Italy}

\date{\today}

\begin{abstract}
We performed a benchmark study on a series of dihydrogen bond complexes
and constructed a set of reference bond distances and interaction energies.
The test set was employed to assess the performance of several
wave-function correlated and density functional theory methods.
We found that second-order correlation methods describe
relatively well the dihydrogen complexes. However, for high accuracy
inclusion of triple contributions is important.
On the other hand, none of the considered density functional 
methods can simultaneously yield
accurate bond lengths and interaction energies. 
However, we found that improved results can be obtained 
by the inclusion of non-local exchange contributions.
\end{abstract}

\maketitle

\section{Introduction}
Noncovalent interactions are of fundamental importance in
a vast number of chemical and physical phenomena.
Thus, they are the subject of numerous computational
studies
\cite{sherrill13,hohenstein12,burns11,thanthiriwatte11,sherrill09,dubecky13,sedlak,hobza13,melichercik13,riley13,riley13_2,zhao07,zhao06,johnson13,otero13,johnson13_2,contreras11,johnson10,grabowski13,noncov_book}
Among others, hydrogen bonds, have a prominent
role in this context, due to their practical and 
historical importance
\cite{hbond_book1,hbond_book2,kollman,zhao12,grabowski11,li11_2,contreras11_2,johnson09,grabowski13_2,fuster11}.

A peculiar type of hydrogen bond is the dihydrogen bond 
in which the bonding occurs between two oppositely
polarized hydrogen atoms. In the most naive physical picture
it can be represented as X--H$^{\delta^-}\cdots^{\delta^+}$H--Y, where X is an 
element less electronegative than H, such as Li, Be, B, Na,
whereas Y is an element more electronegative
than H, such as F, Cl, or CH$_3$.
However, the nature of the bonding cannot be 
simply attributed only to electrostatic effects.
In fact, more complicated quantum effects, such
as exchange and correlation, play a 
prominent role in most cases and they
must be properly taken into account
when an accurate description of the bond is
sought \cite{dihydro_book,dihydro_rev}.
The complex nature of the interactions
beyond dihydrogen bonding reflect the
fact that dihydrogen bonds,
similarly to  conventional hydrogen bonds,
display a strong directionality and
a wide variety of strengths, ranging from few 
tenths of to several tens of kcal/mol,
with no sharp boundary with
dispersion interactions \cite{grabowski04}.

Over the last years, dihydrogen bonding has attracted
great interest, because it has been found to play an important
role for the structure and reactivity of both molecular complexes
and solid-state systems (see for example Refs. 
\citenum{dihydro_book,dihydro_rev} and references therein).
A dihydrogen bond can occur in fact between hydrogen
atoms within a single molecule, or between hydrogens belonging
to different molecules. Thus, it has special relevance 
in many different fields ranging from organic chemistry to
crystal engineering and catalysis.
Moreover, a dihydrogen bond can be viewed as a precursor 
to dehydrogenation reactions \cite{hu04}.

The description of different dihydrogen bonds has been the subject
of numerous theoretical investigations both at the correlated wave-function 
\cite{dihydro_book,dihydro_rev,hayashi05,solimannejad05,alkorta06,solimannejad06,solimannejad06_2,yao11,li11,meng05,filippov06,hugas07,guo13,li13,grabowski13_3} 
and density functional theory (DFT) levels 
\cite{dihydro_book,dihydro_rev,meng05,filippov06,hugas07,guo13,li13,
filippov06_2,zhang11,sandhya12,flener12,yang12}.
These studies analyzed in some details different properties
of a wide number of complexes, computing structures and interaction energies
as well as studying the nature and peculiar characteristics
of this bonding.
However, to date, only few benchmark studies
\cite{dihydro_book,grabowski00} 
have been considered for this important topic.

In this paper we aim to cover this issue and provide a systematic
benchmark investigation of several dihydrogen bond complexes.
Thus, our work has a twofold
goal. First, to provide reference results, from accurate theoretical 
calculations, which can be useful for successive assessment works.
Second, to provide the assessment of different theoretical approaches,
both based on wave-function theory and on DFT, for the 
so constructed benchmark set.

To this end, we firstly consider a representative set of small complexes,
amenable of high-level calculations, including some
typical examples of dihydrogen bonding.
This permits to construct a reliable and methodical test set
which will be very useful as a reference and  
to assess and validate future calculations on 
systems displaying dihydrogen bonds.
At a second step, we perform on the test set
a series of calculations using a wide range of methods
based on wave-function and density functional theory,
in order to understand the expectable accuracy of different 
approaches for the description of the structural
and energetic properties of different complexes.
Finally, we try to correlate the different results
with peculiar characteristics of the different bonds examined.

\section{Computational details}
In our study we considered a set of 32 dihydrogen bond complexes.
The test set was constructed considering the interaction
of different hydrides with several electron donor/acceptor.
Thus, the set can be divided into four subgroups:
\begin{itemize}
\item Complexes formed by \textbf{hydrides of alkali metals}. They
have the general form X--H$\cdots$H--Y, with X=Li, Na and
Y=F, Cl, CN, CCH. 
\item Complexes formed by \textbf{hydrides of elements of group 3A}. This
includes complexes represented by X--H$\cdots$H--Y, with X=B, Al, Ga and
Y=F, Cl, CN, CCH.
\item Linear complexes including \textbf{dihydrides of elements of group 2A}.
They have the general form H--X--H$\cdots$H--Y, with X=Be, Mg and
Y=F, Cl, CN, CCH. 
\item Complexes of \textbf{silane}, i.e.  H$_3$Si--H$\cdots$H--Y, with 
Y=F, Cl, CN, CCH. 
\end{itemize}

For all systems equilibrium structures and interaction energies
were computed using several wave-function correlated methods:
coupled-cluster singles and doubles with perturbative triple
(CCSD(T)) \cite{ccsdt}; quadratic configuration interaction
single and double (QCISD) \cite{qcisd} (also including triple correction
(QCISD(T)) \cite{qcisdt_1,qcisdt_2}); M\o ller-Plesset perturbation
theory \cite{mp} at second order (MP2) \cite{mp2},
average of second and third order (MP2.5) \cite{mp2.5}, 
and fourth order (MP4) \cite{mp4}.
In addition, an energy decomposition analysis was carried on,
based on the symmetry-adapted perturbation theory truncated at
second order in the interaction potential, and at third order in
the monomer fluctuation potential (SAPT2+3) \cite{hohenstein10}.
Finally,
DFT calculations were performed using the exchange-correlation
functionals listed in Table \ref{tab1}.
\begin{table}
\begin{center}
\caption{\label{tab1}Exchange-correlation functionals used in this work. The second column 
indicates the type of the functional: local density approximation (LDA), generalized gradient 
approximation (GGA), meta-GGA, hybrid, meta-GGA hybrid (meta-GGA-H), range-separated hybrid (rs-hybrid).}
\begin{ruledtabular}
\begin{tabular}{llr}
Functional & type & reference \\
\hline
S-VWN      & LDA         & \citenum{slater51,dirac29,vwn} \\
B-P        & GGA         & \citenum{b88,p86} \\
B-LYP      & GGA         & \citenum{b88,lyp} \\
O-LYP      & GGA         & \citenum{lyp,optx} \\
PBE        & GGA         & \citenum{pbe} \\
PBEint     & GGA         & \citenum{pbeint,subroutines} \\
APBE       & GGA         & \citenum{apbe,subroutines} \\
TPSS       & meta-GGA    & \citenum{tpss} \\
revTPSS    & meta-GGA    & \citenum{revtpss} \\
BLOC       & meta-GGA    & \citenum{bloc,loc,bhole,subroutines}  \\
VSXC       & meta-GGA    & \citenum{vsxc} \\
M06-L      & meta-GGA    & \citenum{m06l} \\
B3-LYP     & hybrid      & \citenum{b88,lyp,b3lyp_1,b3lyp_2} \\
BH-LYP     & hybrid      & \citenum{b88,lyp,bhlyp} \\
O3-LYP     & hybrid      & \citenum{lyp,optx,o3lyp} \\
PBE0       & hybrid      & \citenum{pbe,pbe0} \\
hPBEint    & hybrid      & \citenum{pbeint,hpbeint,subroutines} \\
TPSSh      & meta-GGA-H  & \citenum{tpss,tpssh} \\
M06        & meta-GGA-H  & \citenum{m06} \\
M06-HF     & meta-GGA-H  & \citenum{m06,m06hf} \\
CAM-B3LYP  & rs-hybrid   & \citenum{yanai04} \\
LC-BLYP    & rs-hybrid   & \citenum{tawada04} \\
$\omega$B97& rs-hybrid   & \citenum{chai08} \\
$\omega$B97X& rs-hybrid   & \citenum{chai08} \\
\end{tabular}
\end{ruledtabular}
\end{center}
\end{table}

Geometry optimizations with wave-function methods used the aug-cc-pVTZ 
basis set \cite{aug-cc-pvtz1,aug-cc-pvtz2,aug-cc-pvtz3}.
{This basis set was our best compromise between accuracy and the
need to limit the computational cost in order to be able to
carry on all calculations on all systems. 
Test calculations indicated indeed that
this basis set can guarantee an accuracy of $\sim10$ m\AA{}
for the various wave-function calculations.
This result is in agreement with previous studies which evidenced
the appropriateness of such a basis set, stressing the importance
of diffuse functions together with the less prominent role
of the number valence functions \cite{dihydro_book,grabowski00,danovich13}.
}
The final optimized structures were verified to be real minima,
by considering a vibrational analysis at the MP2/aug-cc-pVTZ level
of theory.
{All} interaction energies,
{including DFT ones,} were computed using QCISD(T) optimized geometries.
In fact, QCISD(T) was the higher level of theory for which
we could perform a geometry optimization for all complexes.
Test calculations on the smallest complexes showed anyway that
QCISD(T) results are extremely close to CCSD(T) ones.
{The use of the same geometry for all energy calculations was
considered in order to allow a more homogeneous comparison
between the different methods (i.e. observed differences need
only to be discussed in terms of the different definitions
of the energy in each method). Moreover, the relaxation 
of geometry was found not to modify substantially the observed
trends.}

The energies computed with wave-function methods were
extrapolated to the complete basis set (CBS) limit by considering a 
cubic interpolation formula \cite{halkier98,myrpa} between cc-pVQZ 
\cite{aug-cc-pvtz1,aug-cc-pvtz2,aug-cc-pvtz3} and cc-pV5Z 
\cite{aug-cc-pvtz1,aug-cc-pvtz2,aug-cc-pvtz3} results,
except for CCSD(T) calculations that were extrapolated
to the CBS limit using a focal point analysis \cite{east93,csaszar98}
($\Delta$CCSD(T) procedure) based on CCSD(T)/cc-pVQZ,
MP2/cc-pVQZ, and MP2/CBS results.
{Such energies can be considered accurate within 1\%
or $\sim$0.05 kcal/mol with respect to the true CBS limit \cite{burns14,mackie11,feller11}}
All DFT calculations were performed using the def2-TZVPP
basis set \cite{def2tzvpp1,def2tzvpp2}.
{The choice of this basis set was dictated by the fact that we wanted to test
DFT methods in conditions resembling those used in real applications, 
where usually very large basis sets are not employed. In fact. DFT is seldom used
for benchmarking purposes, but rather as an efficient computational tool
for real applications.} 
All calculations have been corrected for the basis set superposition
error by using a Boys-Bernardi counterpoise correction \cite{cp}.

Calculations were performed using the TURBOMOLE \cite{turbomole}
program package {(DFT methods), the ORCA program 
\cite{orca} (range-separated hybrids),} and the PSI4 code \cite{psi4}
{(wave-function methods)}.

\section{Wave-function calculations}
This section reports the equilibrium H$\cdots$H
bond distance and interaction energy computed with
different wave-function methods. The highest-level
results (QCISD(T) for geometry and CCSD(T) for energies)
are assumed as benchmark values and used as reference to 
assess the performance of all other methods.

\subsection{Equilibrium H$\cdots$H bond distance}
\begin{figure}[b]
\includegraphics[width=\columnwidth]{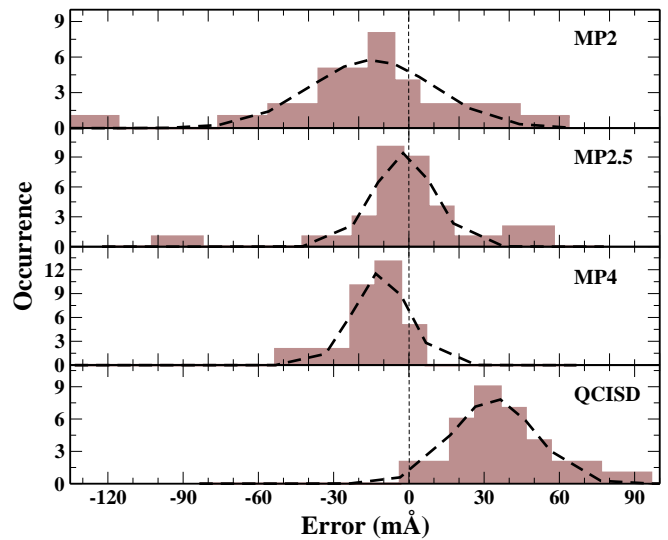}
\caption{\label{fig1} Statistical distribution of the errors on the optimized H$\cdots$H bond 
length for various methods. The dashed lines indicate interpolated Gaussian curves.}
\end{figure}

Tables \ref{tab2} and \ref{tab3}, as well as Fig. \ref{fig1},
report the optimized H$\cdots$H bond
distance for the dihydrogen complexes, as obtained from various
wave-function methods.
\begin{table}
\begin{center}
\caption{\label{tab2}Optimized H$\cdots$H bond distance (\AA) for different dihydrogen bond 
complexes. For each group of systems and the overall set the mean error (ME), mean absolute error 
(MAE), mean absolute relative error (MARE) and the standard deviation with respect to QCISD(T) 
data are reported (in m\AA). [Continues in Table \ref{tab3}].}
\begin{ruledtabular}
\begin{tabular}{lrrrrr}
System & MP2 & MP2.5 & MP4 & QCISD & QCISD(T) \\
\hline
\multicolumn{6}{c}{Hydrides of alkali metals} \\
LiH-HF &  1.3451 &  1.3656 &  1.3502 &  1.3858 &  1.3655 \\
LiH-HCl &  1.2703 &  1.3043 &  1.3217 &  1.3744 &  1.3330 \\
LiH-HCN &  1.6978 &  1.7114 &  1.6990 &  1.7470 &  1.7167 \\
LiH-HCCH &  1.8959 &  1.8974 &  1.8937 &  1.9437 &  1.9061 \\
NaH-HF &  1.2496 &  1.2698 &  1.2532 &  1.2843 &  1.2682 \\
NaH-HCl &  0.9299 &  0.9648 &  1.0236 &  1.1282 &  1.0559 \\
NaH-HCN &  1.5609 &  1.5758 &  1.5545 &  1.6095 &  1.5697 \\
NaH-HCCH &  1.7371 &  1.7536 &  1.7193 &  1.7954 &  1.7446 \\
 &   &  &   &   &  \\
ME & -34.1 & -14.6 & -18.1 & 38.6 &   \\
MAE & 34.1 & 18.8 & 18.1 & 38.6 &   \\
MARE &  2.78\% &  1.57\% &  1.29\% &  2.74\% &  \\
Std.Dev. & 41.1 & 33.0 &  7.2 & 17.7 &   \\
\hline
\multicolumn{6}{c}{Hydrides of group-3A elements} \\
BH-HF &  1.7997 &  1.7998 &  1.7488 &  1.8043 &  1.7490 \\
BH-HCl &  1.8111 &  1.8144 &  1.8100 &  1.8806 &  1.8034 \\
BH-HCN &  2.1461 &  2.1459 &  2.1001 &  2.1562 &  2.1045 \\
BH-HCCH &  2.1436 &  2.1436 &  2.1212 &  2.1574 &  2.1235 \\
AlH-HF &  1.4689 &  1.4877 &  1.4715 &  1.5081 &  1.4855 \\
AlH-HCl &  1.4967 &  1.5181 &  1.5247 &  1.5720 &  1.5345 \\
AlH-HCN &  1.7387 &  1.7432 &  1.7371 &  1.7700 &  1.7461 \\
AlH-HCCH &  1.8336 &  1.8371 &  1.8325 &  1.8725 &  1.8418 \\
GaH-HF &  1.4025 &  1.4221 &  1.4077 &  1.4357 &  1.4220 \\
GaH-HCl &  1.4450 &  1.4730 &  1.4685 &  1.5127 &  1.4846 \\
GaH-HCN &  1.6653 &  1.6931 &  1.6724 &  1.7213 &  1.6956 \\
GaH-HCCH &  1.6802 &  1.7233 &  1.6919 &  1.7615 &  1.7279 \\
 &   &   &   &   &   \\
ME & -7.2 &  6.9 & -11.0 & 36.2 &  \\
MAE & 27.3 & 14.0 & 12.1 & 36.2 &   \\
MARE &  1.61\% &  0.77\% &  0.74\% &  2.07\% &   \\
Std.Dev. & 31.8 & 20.7 & 11.2 & 17.5 &  \\
\end{tabular}
\end{ruledtabular}
\end{center}
\end{table}
\begin{table}
\begin{center}
\caption{\label{tab3}[Continues from Table \ref{tab2}]. Optimized H$\cdots$H bond distance (\AA) 
for different dihydrogen bond complexes. For each group of systems and the overall set the mean 
error (ME), mean absolute error (MAE), mean absolute relative error (MARE) and the standard 
deviation with respect to QCISD(T) data are reported (in m\AA).}
\begin{ruledtabular}
\begin{tabular}{lrrrrr}
System & MP2 & MP2.5 & MP4 & QCISD & QCISD(T) \\
\hline
\multicolumn{6}{c}{Dihydrides of group-2A elements} \\
HBeH-HF &  1.5651 &  1.5782 &  1.5644 &  1.5975 &  1.5718 \\
HBeH-HCl &  1.6761 &  1.6927 &  1.6825 &  1.7504 &  1.7055 \\
HBeH-HCN &  1.9272 &  1.9277 &  1.9238 &  1.9609 &  1.9262 \\
HBeH-HCCH &  2.0570 &  2.0589 &  2.0534 &  2.0872 &  2.0534 \\
HMgH-HF &  1.4486 &  1.4644 &  1.4502 &  1.4812 &  1.4622 \\
HMgH-HCl &  1.5070 &  1.5264 &  1.5318 &  1.5752 &  1.5410 \\
HMgH-HCN &  1.7764 &  1.7787 &  1.7737 &  1.8163 &  1.7840 \\
HMgH-HCCH &  1.8927 &  1.8959 &  1.8892 &  1.9417 &  1.8984 \\
 &   &   &   &   &   \\
ME & -11.5 & -2.4 & -9.2 & 33.5 &  \\
MAE & 12.7 &  6.3 &  9.2 & 33.5 &   \\
MARE &  0.78\% &  0.38\% &  0.55\% &  1.92\% &  \\
Std.Dev. & 13.5 &  7.9 &  6.9 &  8.5 &  \\
\hline
\multicolumn{6}{c}{Silane} \\
SiH$_4$-HF &  1.6953 &  1.6984 &  1.6904 &  1.7324 &  1.7253 \\
SiH$_4$-HCl &  1.7634 &  1.7854 &  1.7850 &  1.8555 &  1.7977 \\
SiH$_4$-HCN &  1.9941 &  2.0003 &  1.9748 &  2.0366 &  1.9893 \\
SiH$_4$-HCCH &  2.0750 &  2.0810 &  2.0620 &  2.1172 &  2.0728 \\
 &  &   &   &  &   \\
ME & -14.3 & -5.0 & -18.2 & 39.2 &  \\
MAE & 17.8 & 14.6 & 18.2 & 39.2 &  \\
MARE &  1.00\% &  0.80\% &  1.00\% &  2.04\% &   \\
Std.Dev. & 20.7 & 17.9 & 11.2 & 22.2 &   \\
\hline
\multicolumn{6}{c}{Overall performance} \\
ME & -15.9 & -2.3 & -13.2 & 36.5 &  \\
MAE & 24.2 & 13.4 & 13.6 & 36.5 &   \\
MARE &  1.62\% &  0.88\% &  0.86\% &  2.19\% &  \\
Std.Dev. & 30.7 & 22.7 &  9.7 & 15.7 &  \\
\end{tabular}
\end{ruledtabular}
\end{center}
\end{table}
Inspection of the data shows that {a proper} inclusion of triple contributions
is very important to achieve good accuracy. In fact,
both MP2.5 and MP4 agree well with reference QCISD(T) calculations,
showing differences that are on average below 1\%.
{Nevertheless, the MP2.5 error distribution is considerably
broader than the MP4 one (see Fig. \ref{fig1}), indicating that the
former method shows limitations for some specific systems.
In more detail, we see that these occur for the BH- complexes and
NaH-HCl. All these complexes are characterized by a relevant
role of the long-range intermolecular forces (induction and/or
dispersion; see later on Table \ref{tab_sapt}). Thus, we
can argue that higher-order correlation contributions are
very important in these cases. 
}
On the contrary, significantly larger errors are found for 
{the second-order MP2 method, which displays}
deviations from the reference larger than 20-30 m\AA.
{A similar performance is also given by the QCISD approach
(as well as by the CCSD method which is almost identical to QCISD).}
In particular, we note that QCISD calculations yield the worst
average results for all groups of complexes, showing always
a marked overestimation of the H$\cdots$H bond length.
For MP2 slightly better results are observed on average. However,
in this case the distribution of the errors is much more erratic,
with some very small errors for some systems and rather large errors
for others (see the standard deviation values in Tabs. \ref{tab2} and
\ref{tab3} and Fig. \ref{fig1}).

\subsection{Interaction energy}
\begin{figure}[b]
\includegraphics[width=\columnwidth]{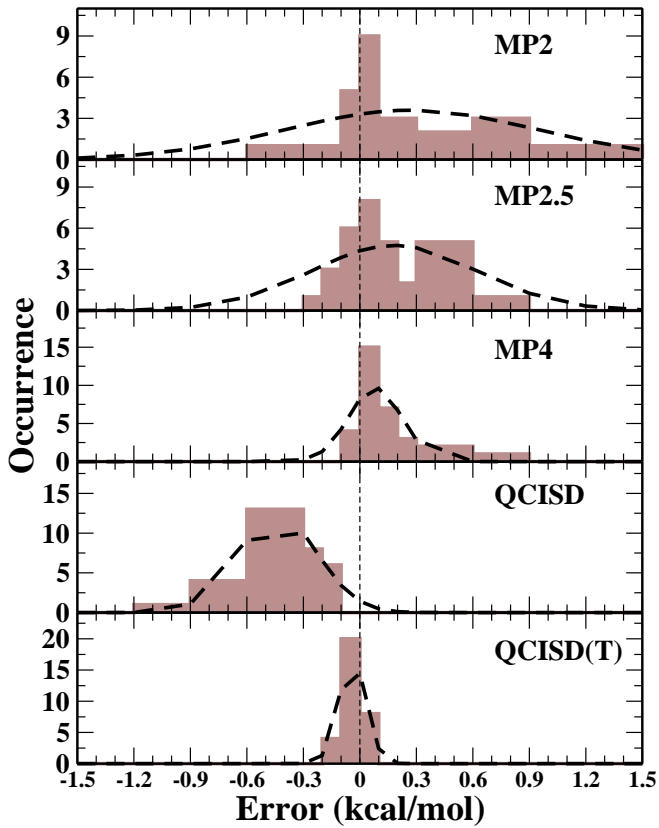}
\caption{\label{fig2} Statistical distribution of the errors on the interaction energy for 
various methods. The dashed lines indicate interpolated Gaussian curves.}
\end{figure}

The interaction energy of the different dihydrogen complexes, calculated
with several wave-function correlated methods, is reported in Tabs.
\ref{tab4} and \ref{tab5} and Fig. \ref{fig2}.
\begin{table*}
\begin{center}
\caption{\label{tab4} Interaction energy (kcal/mol) for the dihydrogen bond complexes. For each 
group of systems and the overall set the mean error (ME), mean absolute error (MAE), mean 
absolute relative error (MARE) and the standard deviation with respect to CCSD(T) data are 
reported. [Continues in Table \ref{tab5}].}
\begin{ruledtabular}
\begin{tabular}{lrrrrrr}
System &    MP2 &    MP2.5 &    MP4 &    QCISD &    QCISD(T) &   CCSD(T) \\
\hline
\multicolumn{7}{c}{Hydrides of alkali metals} \\
LiH-HF &   14.65 &   14.40 &   14.30 &   13.59 &   14.04 &   14.22 \\
LiH-HCl &   13.30 &   12.74 &   12.23 &   11.09 &   11.90 &   11.88 \\
LiH-HCN &    8.70 &    8.70 &    8.63 &    8.14 &    8.40 &    8.46 \\
LiH-HCCH &    4.25 &    4.23 &    4.24 &    3.80 &    4.07 &    4.12 \\
NaH-HF &   16.34 &   15.90 &   15.59 &   14.58 &   15.14 &   15.33 \\
NaH-HCl &   24.77 &   23.59 &   22.05 &   20.39 &   21.46 &   21.38 \\
NaH-HCN &    8.97 &    8.88 &    8.69 &    7.90 &    8.28 &    8.35 \\
NaH-HCCH &    4.09 &    4.00 &    3.95 &    3.26 &    3.66 &    3.73 \\
 &     &     &     &     &     &     \\
ME &    0.95 &    0.62 &    0.28 &   -0.59 &   -0.07 &    \\
MAE &    0.95 &    0.62 &    0.28 &    0.59 &    0.09 &   \\
MARE &    7.56\% &    5.21\% &    2.90\% &    6.27\% &    0.96\% &   \\
Std.Dev. &    1.07 &    0.69 &    0.19 &    0.24 &    0.09 &   \\
\hline
\multicolumn{7}{c}{Hydrides of group-3A elements} \\
BH-HF &    0.20 &    0.22 &    0.48 &    0.28 &    0.52 &    0.52 \\
BH-HCl &    0.10 &    0.05 &    0.20 &   -0.11 &    0.19 &    0.19 \\
BH-HCN &   -0.24 &   -0.19 &   -0.01 &   -0.19 &    0.03 &    0.04 \\
BH-HCCH &   -0.06 &   -0.06 &    0.05 &   -0.10 &    0.05 &    0.05 \\
AlH-HF &    6.15 &    6.02 &    6.06 &    5.56 &    5.90 &    6.07 \\
AlH-HCl &    4.61 &    4.34 &    4.19 &    3.44 &    3.99 &    4.03 \\
AlH-HCN &    3.00 &    2.97 &    3.01 &    2.59 &    2.85 &    2.94 \\
AlH-HCCH &    1.38 &    1.34 &    1.40 &    1.01 &    1.28 &    1.33 \\
GaH-HF &    5.51 &    5.40 &    5.44 &    4.91 &    5.30 &    5.48 \\
GaH-HCl &    4.40 &    4.11 &    3.90 &    3.12 &    3.72 &    3.75 \\
GaH-HCN &    2.54 &    2.51 &    2.55 &    2.10 &    2.39 &    2.46 \\
GaH-HCCH &    0.74 &    0.68 &    0.74 &    0.29 &    0.62 &    0.66 \\
 &     &     &     &     &     &     \\
ME &    0.07 &   -0.01 &    0.04 &   -0.39 &   -0.06 &     \\
MAE &    0.20 &    0.14 &    0.06 &    0.39 &    0.06 &     \\
MARE &   90.31\% &   79.40\% &   14.19\% &  103.00\% &    4.05\% &    \\
Std.Dev. &    0.29 &    0.19 &    0.07 &    0.16 &    0.06 &     \\
\end{tabular}
\end{ruledtabular}
\end{center}
\end{table*}
\begin{table*}
\begin{center}
\caption{\label{tab5} [Continues from Table \ref{tab4}]. Interaction energy (kcal/mol) for the 
dihydrogen bond complexes. For each group of systems and the overall set the mean error (ME), 
mean absolute error (MAE), mean absolute relative error (MARE) and the standard deviation with 
respect to CCSD(T) data are reported.}
\begin{ruledtabular}
\begin{tabular}{lrrrrrr}
System &    MP2 &    MP2.5 &    MP4 &    QCISD &    QCISD(T) &   CCSD(T) \\
\hline
\multicolumn{7}{c}{Dihydrides of group-2A elements} \\
HBeH-HF &    3.43 &    3.39 &    3.44 &    3.14 &    3.36 &    3.42 \\
HBeH-HCl &    2.35 &    2.24 &    2.19 &    1.81 &    2.10 &    2.10 \\
HBeH-HCN &    1.92 &    1.93 &    1.97 &    1.77 &    1.90 &    1.92 \\
HBeH-HCCH &    1.04 &    1.05 &    1.09 &    0.90 &    1.03 &    1.05 \\
HMgH-HF &    7.02 &    6.88 &    6.86 &    6.38 &    6.70 &    6.74 \\
HMgH-HCl &    5.05 &    4.80 &    4.61 &    3.92 &    4.41 &    4.35 \\
HMgH-HCN &    3.64 &    3.63 &    3.63 &    3.27 &    3.48 &    3.48 \\
HMgH-HCCH &    1.82 &    1.81 &    1.84 &    1.50 &    1.72 &    1.73 \\
 &     &    &    &    &     &     \\
ME &    0.19 &    0.12 &    0.11 &   -0.26 &   -0.01 &     \\
MAE &    0.19 &    0.13 &    0.11 &    0.26 &    0.03 &     \\
MARE &    5.40\% &    3.68\% &    3.71\% &    9.83\% &    0.91\% &    \\
Std.Dev. &    0.24 &    0.15 &    0.08 &    0.10 &    0.04 &    \\
\hline
\multicolumn{7}{c}{Silane} \\
SiH4-HF &    1.01 &    0.98 &    1.07 &    0.85 &    1.02 &    1.05 \\
SiH4-HCl &    0.82 &    0.74 &    0.73 &    0.41 &    0.67 &    0.68 \\
SiH4-HCN &    0.61 &    0.60 &    0.65 &    0.47 &    0.61 &    0.62 \\
SiH4-HCCH &    0.34 &    0.32 &    0.37 &    0.19 &    0.34 &    0.35 \\
 &     &     &    &     &     &     \\
ME &    0.02 &   -0.02 &    0.03 &   -0.20 &   -0.02 &     \\
MAE &    0.05 &    0.05 &    0.03 &    0.20 &    0.02 &    \\
MARE &    7.22\% &    6.82\% &    4.95\% &   32.17\% &    2.20\% &   \\
Std.Dev. &    0.08 &    0.05 &    0.01 &    0.05 &    0.01 &   \\
\hline
\multicolumn{7}{c}{Overall performance} \\
ME &    0.31 &    0.18 &    0.11 &   -0.38 &   -0.04 &     \\
MAE &    0.37 &    0.25 &    0.12 &    0.38 &    0.05 &    \\
MARE &   38.01\% &   32.85\% &    7.59\% &   46.67\% &    2.26\% &   \\
Std.Dev. &    0.67 &    0.44 &    0.14 &    0.21 &    0.06 &  \\
\end{tabular}
\end{ruledtabular}
\end{center}
\end{table*}
The results show that, as it may be expected, QCISD(T) calculations
are very close to CCSD(T) ones, with average differences of the order
of 0.06 kcal/mol. This error is close to the expected accuracy of CCSD(T)
calculations \cite{hobza13,feller11,feller06}. Thus, the two methods can be
considered equally accurate from the practical point of view.
{Similarly, almost identical results are found for QCISD and CCSD 
calculations.}

Slightly larger deviations from the CCSD(T) reference are found for 
MP4, which yields a MARE of about 7.5\%
corresponding to a MAE of 0.12 kcal/mol. Overall, MP4 performs very
similarly to QCISD(T) and CCSD(T) for most of the systems. However,
for some of the hydrides of alkali metals (e.g. Na--H$\cdots$H--Cl) 
rather larger errors are found. For these 
systems the relatively poor performance of MP4 shall be traced back
to a worse convergence of the M\o ller-Plesset perturbative expansion,
as indicated by the fact that they show the larger errors also for MP2 and
MP2.5.
{Note that these systems also display a similar behavior for the
geometry errors.}
Furthermore, the MP4 method fails to provide a correct description of
the B--H$\cdots$H-CN complex, which results unbound (by -0.01 kcal/mol) 
at the MP4 level of theory.
We note, however, that this is a particularly difficult case, because 
the reference CCSD(T) interaction energy is only 0.04 kcal/mol.
Thus, small inaccuracies in the CCSD(T) results as well as the
employed QCISD(T) geometry may play a relevant role in this case,
making the comparison uncertain.

{All other methods, i.e. the low-level MP2 method including only
double excitations, as well as the MP2.5 method and QCISD,
including triple corrections, fail to reproduce accurate interaction energies in 
numerous cases.}
In particular, they face limitations to describe
the hydrides of alkali metals, yielding mean absolute errors
larger than 0.6 kcal/mol, and the weakest bonds of the hydrides
of the elements of group 3A. In this latter case, all three methods
predict incorrectly a negative interaction energy.
As a result, the overall performance of MP2, MP2.5, and QCISD
is definitely poorer than the one of MP4 and QCISD(T), with
a mean absolute relative error that is about five time larger.

{
\subsection{Overall performance}
The results of previous subsections indicate that
none of the examined wave-function methods is able to
yield simultaneously reliable binding energies and H$\cdots$H
bond distances, as compared to the reference CCSD(T)/QCISD(T) results,
with the exception of MP4, which performs reasonably well 
for both properties (see Fig. \ref{wft_overall_fig}).
\begin{figure}
\includegraphics[width=\columnwidth]{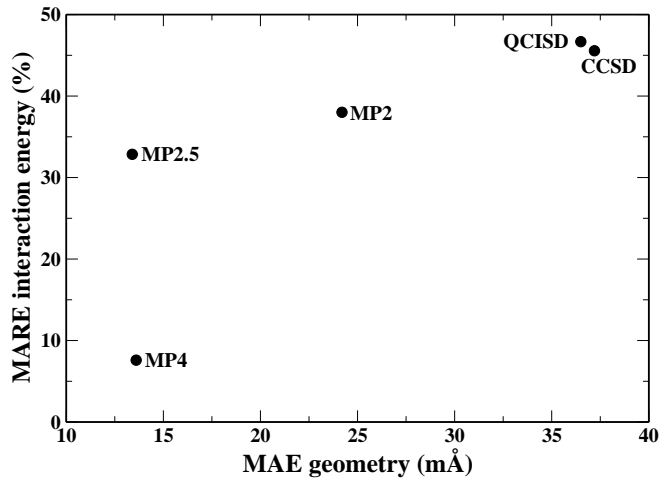}
\caption{\label{wft_overall_fig} Mean absolute error (MAE) for H$\cdots$H bond length versus the mean absolute relative error (MARE) on interaction energies for several wave-function methods. The most accurate methods shall occupy the left bottom corner of the plot.}
\end{figure}
Nevertheless, as we mentioned before, MP4 suffers from
its inability to describe certain systems, for which it shows
errors much above its average. This fact, together with the
relatively high computational cost of the MP4 method, 
contributes to penalize MP4 as a method of choice in the study of
dihydrogen interactions and suggests that, when high
accuracy is sought, QCISD(T) calculations may be
employed instead.

Concerning other, cheaper methods we remark once more
that all display several limitations for
the calculation of binding energies
and/or geometries. However, when computational effort is an issue,
the MP2 (or even better the MP2.5) method appears
to be the best compromise to achieve reasonable accuracy 
with a moderate effort.
We remark in particular that the MP2.5 method is in fact 
able to yield results comparable with MP4 for many systems.
However, it shows limitations for some specific systems
which are more strongly characterized by long-range interactions.
}

\subsection{Energy decomposition analysis}
To understand better the nature of the bonding in the
different complexes we performed an energy decomposition
analysis via SAPT2+3 calculations. The results of
this analysis are listed in Table \ref{tab_sapt}, where the 
electrostatic, exchange, induction and dispersion contributions to the
interaction energy are reported.
In addition we report, in analogy with Ref. \citenum{singh09}, 
the relative weight of each component,
defined as $w=|E_i|/\sum_i|E_i|$, with $E_i$ denoting
the different interaction energy contributions.
\begin{table*}
\begin{center}
\caption{\label{tab_sapt}Different components of the SPAT2+3 energy (kcal/mol) for the dihydrogen bond complexes and, in parenthesis, their relative weight $w$ in \% (see text). }
\begin{ruledtabular}
\begin{tabular}{lrrrrrrrrr}
System & \multicolumn{2}{c}{Electrostatic} & \multicolumn{2}{c}{Exchange} &
\multicolumn{2}{c}{Induction} & \multicolumn{2}{c}{Dispersion} & Total \\
\hline
LiH-HF &   18.89 & (34\%) & -20.54 & (37\%) &  11.38 & (20\%) & 4.87 & ( 9\%) & 14.60 \\
LiH-HCl &  19.11 & (26\%) & -30.36 & (41\%) &  17.09 & (23\%) & 7.37 & (10\%) & 13.21 \\
LiH-HCN &  13.09 & (39\%) & -12.35 & (36\%) &   5.25 & (15\%) & 3.22 & ( 9\%) &  9.21 \\
LiH-HCCH &  6.84 & (35\%) &  -7.52 & (39\%) &   2.76 & (14\%) & 2.37 & (12\%) &  4.45 \\
NaH-HF &   22.37 & (30\%) & -28.68 & (39\%) &  16.78 & (23\%) & 6.38 & ( 9\%) & 16.85 \\
NaH-HCl &  27.18 & (18\%) & -60.05 & (40\%) &  47.02 & (32\%) &14.48 & (10\%) & 28.63 \\
NaH-HCN &  16.83 & (33\%) & -20.17 & (40\%) &   8.67 & (17\%) & 4.68 & ( 9\%) & 10.01 \\
NaH-HCCH &  9.35 & (31\%) & -13.00 & (43\%) &   4.65 & (15\%) & 3.50 & (11\%) &  4.50 \\
BH-HF &     0.38 &  (5\%) &  -3.26 & (47\%) &   1.86 & (27\%) & 1.48 & (21\%) &  0.46 \\
BH-HCl &    0.82 & (10\%) &  -4.16 & (48\%) &   1.65 & (19\%) & 1.98 & (23\%) &  0.29 \\
BH-HCN &    0.04 &  (1\%) &  -1.62 & (49\%) &   0.69 & (21\%) & 0.98 & (29\%) &  0.08 \\
BH-HCCH &   0.29 &  (9\%) &  -1.58 & (48\%) &   0.40 & (12\%) & 1.00 & (31\%) &  0.11 \\
AlH-HF &    8.22 & (29\%) & -11.32 & (39\%) &   5.91 & (21\%) & 3.38 & (12\%) &  6.19 \\
AlH-HCl &   7.81 & (24\%) & -13.99 & (43\%) &   6.24 & (19\%) & 4.41 & (14\%) &  4.46 \\
AlH-HCN &   5.84 & (30\%) &  -8.15 & (42\%) &   3.02 & (16\%) & 2.31 & (12\%) &  3.02 \\
AlH-HCCH &  3.88 & (27\%) &  -6.29 & (44\%) &   1.70 & (12\%) & 2.31 & (16\%) &  1.60 \\
GaH-HF &    8.65 & (26\%) & -13.94 & (41\%) &   7.18 & (21\%) & 3.95 & (12\%) &  5.85 \\
GaH-HCl &   8.37 & (22\%) & -16.45 & (44\%) &   7.45 & (20\%) & 5.02 & (13\%) &  4.39 \\
GaH-HCN &   6.20 & (28\%) &  -9.69 & (43\%) &   3.55 & (16\%) & 3.03 & (13\%) &  3.08 \\
GaH-HCCH &  4.80 & (25\%) &  -9.01 & (47\%) &   2.32 & (12\%) & 2.96 & (16\%) &  1.06 \\
HBeH-HF &   4.64 & (27\%) &  -6.88 & (40\%) &   3.45 & (20\%) & 2.32 & (13\%) &  3.53 \\
HBeH-HCl &  3.80 & (24\%) &  -6.70 & (43\%) &   2.65 & (17\%) & 2.58 & (16\%) &  2.32 \\
HBeH-HCN &  2.89 & (32\%) &  -3.48 & (38\%) &   1.26 & (14\%) & 1.49 & (16\%) &  2.16 \\
HBeH-HCCH & 1.73 & (29\%) &  -2.40 & (40\%) &   0.61 & (10\%) & 1.23 & (21\%) &  1.17 \\
HMgH-HF &   9.45 & (30\%) & -12.46 & (39\%) &   6.49 & (20\%) & 3.51 & (11\%) &  6.99 \\
HMgH-HCl &  8.40 & (25\%) & -14.13 & (43\%) &   6.36 & (19\%) & 4.33 & (13\%) &  4.96 \\
HMgH-HCN &  6.29 & (33\%) &  -7.48 & (39\%) &   2.81 & (15\%) & 2.44 & (13\%) &  4.06 \\
HMgH-HCCH & 3.88 & (30\%) &  -5.47 & (42\%) &   1.55 & (12\%) & 2.06 & (16\%) &  2.02 \\
SiH4-HF &   0.88 & (11\%) &  -3.39 & (43\%) &   1.99 & (25\%) & 1.61 & (20\%) &  1.10 \\
SiH4-HCl &  1.13 & (13\%) &  -4.08 & (45\%) &   1.69 & (19\%) & 2.11 & (23\%) &  0.84 \\
SiH4-HCN &  0.75 & (15\%) &  -2.20 & (43\%) &   0.92 & (18\%) & 1.29 & (25\%) &  0.77 \\
SiH4-HCCH & 0.55 & (14\%) &  -1.76 & (45\%) &   0.47 & (12\%) & 1.18 & (30\%) &  0.43 \\
\end{tabular}
\end{ruledtabular}
\end{center}
\end{table*}
We see that for most systems the energy decomposition
describes a bonding behavior quite similar
to conventional hydrogen bonds \cite{hoja14},
despite for the present dihydrogen bonds we observe in general
a slightly more important role of electrostatic and induction terms
and a reduced influence of the dispersion contributions.
In fact in all cases the largest component of the
interaction energy is the exchange one, which
weights about 40\% and is repulsive. In most cases
the second largest contribution is given by the attractive 
electrostatic interaction, which has a weight
of 25-30\%. Finally, the induction and dispersion
terms provide further, but less important, contributions
to the interaction energy, having on average weights
of about 20\% and 10\%, respectively.

We note, however, that for some systems, 
especially in the set the of hydrides of 
group-3A elements and the complexes of silane, the
importance of dispersion interactions is much larger,
being eventually the second contribution, after exchange,
to the total interaction energy. These systems shall thus be regarded
as laying at the boundary between dihydrogen bond and 
dispersion complexes. This explains why most of these systems
display very low interaction energies and consequently can
be accurately described only by the higher level approaches.

\section{Density functional theory calculations}
In this section we report an assessment of some popular
DFT functionals for the description of the equilibrium 
geometry and the interaction energy of the dihydrogen complexes
studied in the previous section.
We remark that, due to the huge number of existing XC functionals, 
this study is not intended as an exhaustive investigation, but rather aims to
provide a general feeling of the performance that can be expected 
from DFT.
In particular, we did not considered Van der Waals corrected functionals.
In fact, an analysis of the many different existing techniques
for the treatment of Van der Waals forces in the DFT framework
requires a large effort and deserves a separate investigation
(see for example Refs. \citenum{burns11,arey09,hujo11,dilabio13}).
Moreover, the analysis of the results concerning complexes 
with very different energies and bonding nature 
will be very complicated and deserves separate studies.
For these reasons we removed from the test set considered in the
following analysis those complexes where
dispersion interactions are particularly relevant, which are the
ones that have in Table \ref{tab_sapt} the weight of dispersion 
contributions exceeding 20\% (i.e. all BH- and silane complexes
as well as HBeH--HCCH; note that for most of these systems dispersion is
also the second biggest contribution in the interaction energy).

\subsection{Equilibrium H$\cdots$H bond distance}
\begin{figure}[b]
\includegraphics[width=\columnwidth]{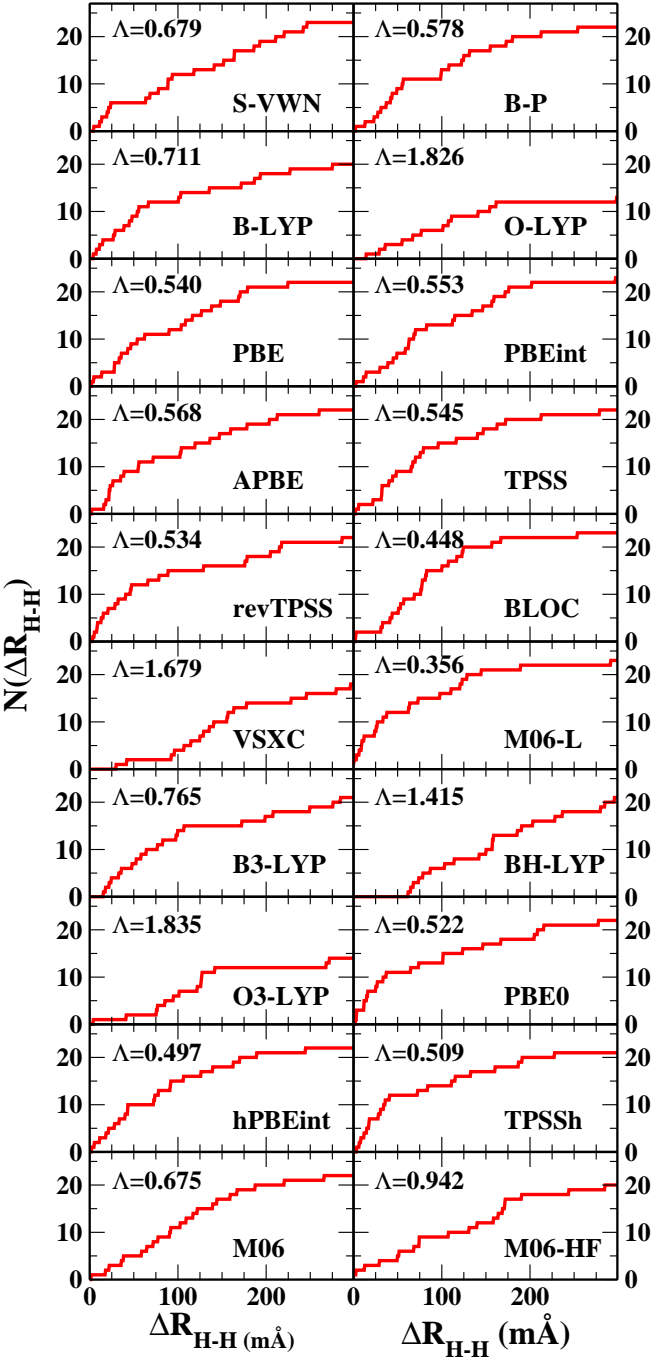}
\caption{\label{fig4} Cumulative number $N(\Delta R_{H-H})$ of systems with
  error lower or equal than $\Delta R_{H-H}$ for different XC functionals. The integrated error $\Lambda$ is also reported (see text for details).}
\end{figure}

Analyzing the performance of DFT functionals for the H$\cdots$H bond distance
we found that the distribution of the errors is spread over a quite large range, 
covering an interval of about $\pm$200 m\AA{} for all the functionals with
even larger errors for some systems.
Moreover, a marked tendency towards the overestimation of the bond distance
is observable in general. 
This situation implies that the performance of different functionals can not be measured by
computing, as usual, mean (absolute) (relative) errors, because for such a broad distribution of data
the average will contain little information. Thus, we prefer to report in Fig. \ref{fig4}, for 
each functional, the histogram of the cumulative number of systems with error below a certain value.
That is, for each functional we define $\Delta R_{H-H}^i$ the absolute error on the H$\cdots$H 
bond length
for the $i$-th complex and  we consider the quantity
\begin{equation}
N(\Delta R_{H-H}) = \sum_{i=1}^{23}f(\Delta R_{H-H}^i;\Delta R_{H-H})\ ,
\end{equation}
where $f(\Delta R_{H-H}^i;\Delta R_{H-H})=1$ if $\Delta R_{H-H}^i\leq\Delta R_{H-H}$ and 0
otherwise.
Furthermore, to obtain a more quantitative evaluation of the performance of different
functionals we consider the indicator $\Lambda=(1-\epsilon)/\epsilon$, where
$\epsilon$ is the integrated error
\begin{equation}
\epsilon = \frac{1}{23R_{max}}\int_0^{R_{max}}N(x)dx\ ,
\end{equation}
with $R_{max}$ being fixed to 300 m\AA. Given that $R_{max}$ is chosen
such that $N(R_{max}) = N_{max}$, the 
indicator $\Lambda$ shows how fast the function $N$ grows to its maximum 
(all curves in Fig. \ref{fig4}
are roughly fitted by  
$N(x)\propto x^{\Lambda}$). 
Therefore, for a perfect functional we would have
$\Lambda=0$, whereas for a very poor functional $\Lambda\rightarrow \infty$.

The plots of Fig. \ref{fig4} show that most functionals perform rather poorly for
the H$\cdots$H equilibrium distance. In fact, in most cases errors below 25 m\AA{}
are obtained only for few systems and even errors below
100 m\AA{} are not very common. According to our analysis the
best performance is given by the M06-L functional ($\Lambda=0.356$), 
which yields 60\% of the complexes with an error less than 50 m\AA. 
Relatively good results are obtained also from 
PBE, PBEint, and APBE among the GGAs,
BLOC among the meta-GGAs, and hPBEint among the hybrids.
Very poor results are given, on the other hand, by O-LYP, VSXC, BH-LYP, 
O3-LYP, and M06-HF, that all make worse than the simple local density 
approximation. Note that also the popular B3-LYP functional displays a rather
disappointing behavior being similar with S-VWN.
Indeed, the inclusion of small fraction of Hartree-Fock 
exchange into the hybrids seems to bring in general 
a slight improvement of the performance, whereas functionals 
including a large amount of Hartree-Fock exchange
display in general poor results (see also later on).

\subsection{Interaction energy}
The mean absolute (relative) errors on the interaction energies computed with different 
XC functionals are reported in Table \ref{tab7}. 
\begin{table*}
\begin{center}
\caption{\label{tab7} Mean absolute errors (kcal/mol) and mean absolute relative errors (in parenthesis) on the interaction energy of different groups of dihydrogen bond complexes. The overall errors are also reported in the last column.}
\begin{ruledtabular}
\begin{tabular}{lrrcrrcrrcrr}
Functional & \multicolumn{2}{c}{alkali metals} & $\; \;$ & \multicolumn{2}{c}{group 3A} & $\; \;$ & \multicolumn{2}{c}{Group 2A} & $\; \;$ & \multicolumn{2}{c}{overall} \\ 
\hline
\multicolumn{12}{c}{LDA/GGA functionals} \\
S-VWN     & 3.45 & (35.35\%) & & 3.13 &(123.21\%) & & 2.08 & (65.89\%) & & 2.89 & (76.75\%) \\
B-P       & 0.61 &  (7.87\%) & & 0.61 & (22.24\%) & & 0.36 & (15.74\%) & & 0.53 & (15.56\%) \\ 
B-LYP     & 1.13 & (14.60\%) & & 0.51 & (24.50\%) & & 0.48 & (19.72\%) & & 0.70 & (19.80\%) \\
O-LYP     & 2.52 & (31.71\%) & & 1.38 & (75.37\%) & & 1.55 & (55.45\%) & & 1.80 & (55.02\%) \\
PBE       & 0.52 &  (3.63\%) & & 0.95 & (29.51\%) & & 0.35 & ( 9.41\%) & & 0.62 & (14.80\%) \\
PBEint    & 0.70 &  (4.99\%) & & 0.91 & (27.06\%) & & 0.38 & (10.76\%) & & 0.67 & (14.78\%) \\
APBE      & 0.41 &  (3.77\%) & & 0.77 & (22.27\%) & & 0.27 &  (7.55\%) & & 0.49 & (11.64\%) \\
\hline
\multicolumn{12}{c}{meta-GGA functionals} \\
TPSS      & 0.64 &  (4.84\%) & & 0.63 & (17.90\%) & & 0.28 & ( 8.29\%) & & 0.52 & (10.65\%) \\
revTPSS   & 0.16 &  (1.98\%) & & 0.44 & (12.50\%) & & 0.20 & ( 6.99\%) & & 0.27 & ( 7.37\%) \\
BLOC      & 0.97 &  (9.38\%) & & 0.92 & (30.99\%) & & 0.44 & (11.46\%) & & 0.78 & (17.83\%) \\
VSXC      & 0.50 &  (4.00\%) & & 0.29 & (10.72\%) & & 0.17 &  (5.67\%) & & 0.32 & ( 6.95\%) \\
M06-L     & 1.07 & (10.84\%) & & 0.62 & (19.98\%) & & 0.28 & ( 9.70\%) & & 0.65 & (13.77\%) \\
\hline
\multicolumn{12}{c}{hybrid functionals} \\
B3-LYP    & 0.50 & (7.16\%) & & 0.36 & (17.11\%) & & 0.30 & (13.03\%) & & 0.39 & (12.62\%) \\
BH-LYP    & 0.24 & (3.07\%) & & 0.25 & (12.72\%) & & 0.20 & ( 8.40\%) & & 0.23 & ( 8.25\%) \\
O3-LYP    & 1.83 &(23.58\%) & & 1.08 & (59.95\%) & & 1.22 & (44.14\%) & & 1.37 & (43.25\%) \\
PBE0      & 0.72 & (5.92\%) & & 0.51 & (14.40\%) & & 0.20 &  (5.42\%) & & 0.48 & ( 8.81\%) \\
hPBEint   & 0.80 & (6.18\%) & & 0.64 & (18.50\%) & & 0.28 & ( 7.96\%) & & 0.57 & (11.18\%) \\
\hline
\multicolumn{12}{c}{hybrid meta-GGA functionals} \\
TPPSh     & 0.66 & (5.20\%) & & 0.50 & (13.72\%) & & 0.24 & ( 7.62\%) & & 0.47 & ( 9.04\%) \\
M06       & 0.72 & (8.06\%) & & 0.60 & (22.27\%) & & 0.14 &  (3.58\%) & & 0.49 & (11.74\%) \\
M06-HF    & 0.43 & (5.00\%) & & 0.56 & (31.18\%) & & 0.62 & (23.07\%) & & 0.54 & (20.20\%) \\
\end{tabular}
\end{ruledtabular}
\end{center}
\end{table*}
The overall performance of all functionals is in line with that
obtained for conventional hydrogen bonds \cite{hb06},
with average errors mostly included in the range 0.3-0.5 kcal/mol.
The best functionals in Table \ref{tab7} turn out to be BH-LYP
(MAE=0.23 kcal/mol, MARE=8.3\%) and revTPSS (MAE=0.27 kcal/mol,
MARE=7.4\%); amongst the GGAs the best performance is 
shown by APBE with a MAE=0.49 kcal/mol (MARE=11.6\%). 
Nevertheless, all functionals, but S-VWN, O-LYP and O3-LYP,
perform quite similarly on average.
Slightly larger differences are observed considering individual groups
of complexes. However, a clear trend cannot be established.
Nevertheless, we can note that in general meta-GGA functionals
yield the best results and the most uniform description of different
systems also among different classes. The inclusion of 
Hartree-Fock exchange in hybrids, appears to slightly improve
the description of the dihydrides of group 2A and 3A elements,
whereas it yields a little worsening for the complexes of the
alkali metals.

To try to rationalize better this behavior we
{consider a density analysis of some representative
complexes. Thus, in the upper panel of Fig. \ref{denspar_fig}
we report a plot of the reduced gradient 
$s=|\nabla n|/[(4(3\pi^2)^{2/3}n^{4/3}]$ as a function of the
electron density ($n$) times the second eigenvalue of the electron-density
Hessian ($\lambda_2$). This is the NCI
indicator \cite{nci,nci2} which is able to characterize different
kinds of non-covalent interactions. Inspection of the plot
shows that all the systems are characterized
by a clear hydrogen-like bonding pattern, although
with varying strengths, without significant
van der Waals signatures even for the weakest complexes
(e.g. GaH-HCCH). This fact confirms that our selection
of systems with (almost) no dispersion character,
performed on the basis of the SAPT energy decomposition
is in fact efficient. Moreover, it suggests that
a deeper analysis can be brought on on the basis of some
semilocal density indicators.
\begin{figure}
\includegraphics[width=0.9\columnwidth]{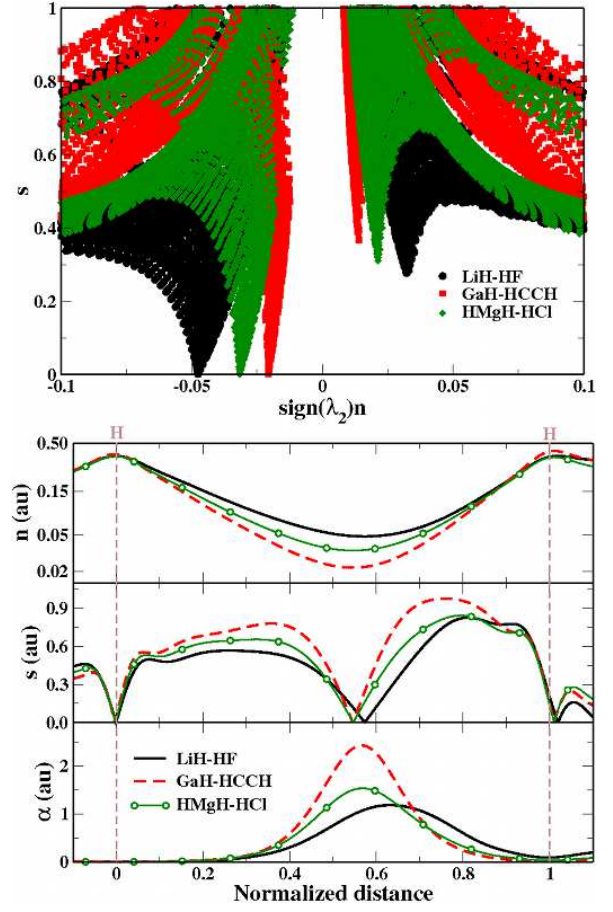}
\caption{\label{denspar_fig} Plot of the NCI indicator \cite{nci,nci2} (upper panel) and of several density parameters as functions of the normalized bond distance (lower panel) for the LiH-HF, LiH-HCCH, and BH-HCCH complexes. See text for details.}
\end{figure}

In the lower panel of Fig. \ref{denspar_fig} we report
}
the plot of some 
important density indicators in the bond region of three
exemplary complexes. Namely, we plot the electron density,
the reduced gradient 
which denotes regions where the density is slowly- or rapidly-varying,
and the meta-GGA ingredient $\alpha=(\tau-\tau^W)/\tau^{TF}$
where $\tau$ is the positive defined kinetic energy density,
$\tau^W=|\nabla n|^2/(8n)$ is the von Weizs\"acker kinetic
energy density, and $\tau^{TF}=(3/10)(3\pi^2)^{2/3}n^{5/3}$ is
the Thomas-Fermi kinetic energy. The latter distinguishes between
iso-orbital regions and slowly-varying regions.
In the plot, for each complex, the distance is normalized
to the H$\cdots$H bond distance, so that the curves are all
comparable.

The figure indicates how difficult may be for a semilocal
DFT functional to differentiate between various complexes.
In fact, despite the three complexes considered for the plots
have interaction energies that vary from 14.22 (LiH-HF) 
to 0.66 kcal/mol (GaH-HCCH), they display only minor differences 
as to what concerns the density and the reduced gradient $s$
in the bond region. In particular, whereas the density
shows a weak trend with the interaction strength 
(complexes with strongest binding have a slightly larger
density in the bond) the reduced gradient $s$ is very similar
in all cases. On the other hand, important differences 
between the various complexes can
be observed by inspecting the meta-GGA indicator $\alpha$.
This helps to explain the better performance of meta-GGA functionals
with respect to GGA ones in terms of their
superior ability to discriminate the nature of the different
bonding patterns.

Additionally, Fig. \ref{denspar_fig} shows that
for all complexes the bonding region is
fundamentally a slowly-varying density region,
since $s\lessapprox0.8$ there. This explains
the failure of the functionals based on the OPTX exchange \cite{optx}
(e.g. O-LYP, O3-LYP) 
which even fail to recover the local density approximation limit.
Nevertheless, it must be noted that the proper
slowly-varying density limit is only observed
in the strongest dihydrogen complexes
(e.g. LiH-HF) where $\alpha\approx 1$ in the bond. 
For the complexes displaying a weakest interaction
instead $\alpha$ is quite larger
indicating that the bonding region 
is an evanescent region for the density
characterized by the contribution of many orbitals
(otherwise $\alpha$ would be zero as in 1 or 2 electron systems).
This situation resembles the interaction of two
closed-shell atoms, and cannot be easily described at
the semilocal level of theory.
Thus, we have an additional element to explain the
superiority of meta-GGA functionals in this context
(group-2A and especially group-3A complexes).
Furthermore, this finding strongly helps
to rationalize the fact that the inclusion
of Hartree-Fock exchange generally improves the
performance of the functionals for the description
of interaction energies (see also next subsection), 
especially in the case
of complexes of group-3A and group-2A elements.
In fact, the inclusion of non local exchange contributions
is likely to improve the description of non local interactions
between the two weakly overlapping densities.

\subsection{Hybrid functionals}
\label{hyb_sec}
\begin{figure}[b]
\includegraphics[width=\columnwidth]{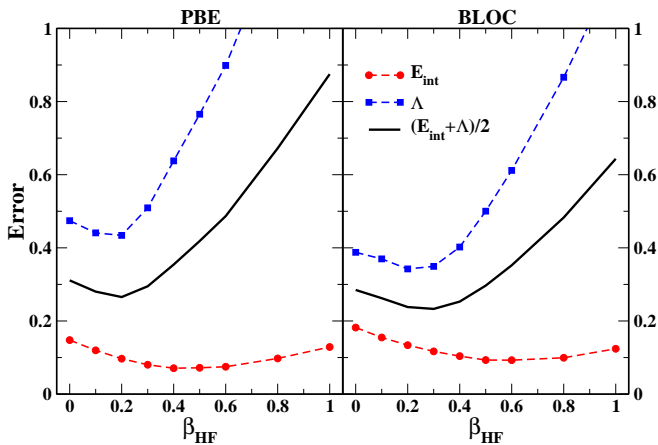}
\caption{\label{fig_hyb} Indicator for geometry errors $\Lambda$ and the mean absolute relative error on interaction energies for different values of the fraction of Hartree-Fock exchange in functionals of the form given by Eq. (\ref{hybeq}) with DFT=PBE (left) and DFT=BLOC (right). In each panel we also report the global performance computed as the average between $\Lambda$ and the MAE.}
\end{figure}

To investigate in some more detail the
role of nonlocal Hartree-Fock exchange in hybrid functionals
we consider in this subsection a couple of model
hybrid XC functionals of the form
\begin{equation}\label{hybeq}
E_{xc}^\mathrm{hybrid} = \beta_{HF}E_x^{HF} + (1-\beta_{HF})E_x^{DFT} + E_c^{DFT}\ ,
\end{equation}
where $\beta_{HF}$ is a parameter, 
$E_x^{HF}$ is the Hartree-Fock exchange energy, $E_x^{DFT}$
is some semilocal DFT exchange functional, and $E_c$ is
a DFT correlation functional. A similar model was used in
Refs. \citenum{fdeene,fdect}.
In this work we consider DFT=PBE, BLOC and we compute
the MAE on interaction energies as well the value of the
indicator $\Lambda$ while $\beta_{HF}$ is varied between 0 and 1.
The results of these calculations are reported in Fig.
\ref{fig_hyb}.

The plot shows that for both PBE and BLOC a similar behavior
is obtained (which is common also to
other functionals not shown),
with a decrease of the errors for rather small
fractions of Hartree-Fock exchange and a 
worsening of the performance when a larger amount of nonlocal
exchange is considered. This trend is similar for
both the geometry and interaction energy errors.
However, for the former case the benefits are observed only 
for small fractions of Hartree-Fock exchange, while
a significant increase of the errors is obtained for $\beta_{HF}>0.4$;
For interaction energies instead a larger fraction of
Hartree-Fock exchange is required for better performance
and no dramatic worsening of the results is
achieved even for $\beta_{HF}=1$.
Thus, all in all we can estimate both functionals
to have a ``best'' average performance at a 
moderately small fraction of Hartree-Fock exchange mixing,
i.e. at about 20\% 
(we denote these ``best'' hybrids hPBE with $\beta_{HF}=0.20$
and hBLOC with $\beta_{HF}=0.25$).

We note however, that the results shown in Fig. \ref{fig_hyb}
are only an average over the various systems that
display in general very different behaviors.
For example, for interaction energies 
the inclusion of Hartree-Fock exchange has 
a quite different effect on complexes of the
alkali metals than on weaker dihydrogen complexes.
In the former case in fact GGA functionals
generally overestimate the interaction energy and
the addition of Hartree-Fock exchange further
increases this overestimation. Thus, the error
usually increases with $\beta_{HF}$. On the other hand,
for complexes of the elements of group 3A, the
GGA functionals mostly provide an overestimation
of the interaction energy but the inclusion
of exact exchange reduces it, so that small
errors are generally obtained at rather
large values of $\beta_{HF}$. Finally,
a mixture of these two trends is observable for
complexes of the group-2A elements.
Therefore, although the inclusion of
a moderate fraction of Hartree-Fock exchange
can be positive for DFT calculations, it
must be kept in mind that 
 a good balance between all the effects
and for different systems, is difficult to achieve.
Thus, caution must be taken before extrapolating
general conclusions to individual cases.

{In consideration of the last comments, we complete this
section by reporting in Tab. \ref{rs_tab} 
the mean absolute relative errors
for interaction energies as obtained by several
range-separated hybrid functionals.
\begin{table*}
\begin{center}
\caption{\label{rs_tab} Mean absolute errors (kcal/mol) and mean absolute relative errors (in parenthesis) on the interaction energy of different groups of dihydrogen bond complexes as computed with everal range-separated DFT functionals. The overall errors are also reported in the last column.}
\begin{ruledtabular}
\begin{tabular}{lrrcrrcrrcrr}
Functional & \multicolumn{2}{c}{alkali metals} & $\; \;$ & \multicolumn{2}{c}{group 3A} & $\; \;$ & \multicolumn{2}{c}{Group 2A} & $\; \;$ & \multicolumn{2}{c}{overall} \\ 
\hline
CAM-B3LYP    & 0.28 & (3.35\%) && 0.33 & (12.82\%) && 0.24 & (9.28\%) && 0.29 & (8.45\%) \\
LC-BLYP      & 0.72 & (6.96\%) && 0.99 & (33.14\%) && 0.59 & (17.21\%)&& 0.77 & (19.18\%) \\
$\omega$B97  & 0.83 & (10.04\%)&& 0.49 & (19.31\%) && 0.48 & (15.98\%)&& 0.61 & (15.07\%) \\
$\omega$B97X & 0.77 & (9.45\%) && 0.50 & (20.57\%) && 0.45 & (15.05\%)&& 0.58 & (15.02\%) \\
\end{tabular}
\end{ruledtabular}
\end{center}
\end{table*}
The table shows that range-separated hybrid functionals perform
generally very well for the complexes under exam. However,
they bring no clear advantage with respect to global hybrid
functionals (CAM-B3LYP is a little better that B3-LYP but worst than BH-LYP;
all other functionals are slightly worst that all global hybrids except
O3-LYP). Thus, the separation between short- and long-range exchange
terms does not appear to be a crucial factor in the treatment
of dihydrogen bonds.

On the other hand, we saw that the functionals incorporating a larger
amount of exact exchange perform better than others for
the description of interaction energies. This issue can be possibly related,
to a smaller delocalization error of these functionals.
}

\subsection{Overall performance}
\begin{figure}[b]
\includegraphics[width=\columnwidth]{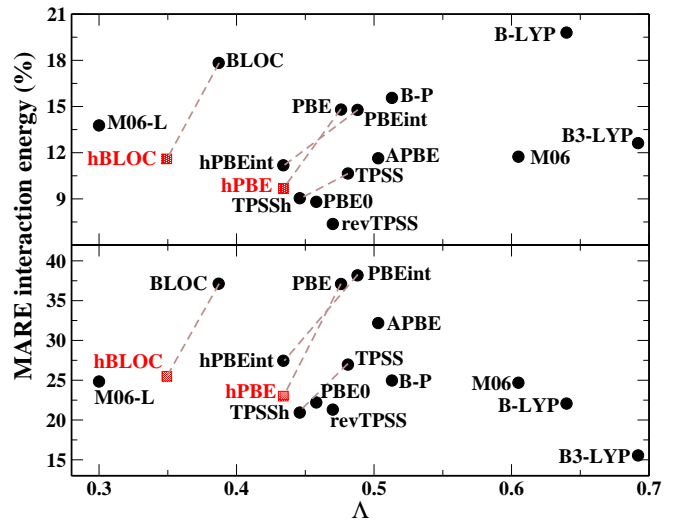}
\caption{\label{fig5} Indicator for geometry errors $\Lambda$ versus the mean absolute relative 
error (MARE) on interaction energies for selected DFT functionals. The top panel report MAREs computed with QCISD(T) geometries; the bottom panel reports the MAREs computed using relaxed geometries. In each panel the most accurate methods shall occupy the left bottom corner of the plot.}
\end{figure}

As we saw a DFT functional
can produce results of different quality 
when the interaction energy or the accuracy of 
the description of the 
H$\cdots$H bonds are considered.
This is an important issue because in practical applications
both the structure and the interaction energy must be
accurately described. Therefore, a well balanced description must
be preferred to a situation where one property is described very well
but the other is not. 
This assessment is presented in Fig. \ref{fig5}, where we report
for selected functionals the value of the indicator $\Lambda$ for 
the geometry versus the mean absolute relative error on the interaction 
energies.
{Note that in the figure, for interaction energy,
both MAREs obtained using QCISD(T) reference
geometries (top panel) and MAREs obtained relaxed DFT geometries (bottom
panel) are considered. The former are in fact more closely related
to the discussion of previous sections, whereas the latter 
are more appropriate for an assessment of practical calculations where the geometry 
and the energy are likely calculated at the same level of theory.
Nevertheless, both cases show very similar trends, while the most evident difference
is that with relaxed DFT geometries the MARE on interaction energy is 
generally increased.}
The most accurate functionals are thus located in the bottom left corner of the plot,
whereas the top right corner will host the worst performing functionals. 
We note that, in this case, the MARE gives a more realistic 
statistical assessment of a given functional than the MAE, 
because the interaction energies of the benchmark systems 
span a considerable range of interaction energies going
from 0.52 kcal/mol 21.38 kcal/mol.

Inspection of the figure shows that the overall performance of DFT
functionals is quite erratic. Nevertheless, there 
exist a group of functionals, including the meta-GGAs revTPSS and M06-L
as well as the hybrids TPSSh, PBE0, and hPBEint, which
perform all quite well, despite none of them can be simultaneously
well accurate for both geometries and energies.
We can rate these functionals as the most reliable for 
applications on dihydrogen bonds.
On the other hand, several functionals, 
mainly GGAs such as PBE and  APBE,
lay in the central part of the figure, showing that they
display a moderate accuracy for both geometries and
interaction energies. This result seems to contrast
with the fact that they are instead quite
good for conventional hydrogen bonds
\cite{mukappa}. However, the results of previous
sections indicate that the overall 
performance of these functionals
is penalized by their inability to describe some
cases (typically the weakest bonds), whereas
they can be a good choice for complexes
where a larger overlap of the fragment 
densities is present.
We remark finally, that in general the performance
of functionals can be improved by the inclusion
of a small fraction of Hartree-Fock exchange in a
hybrid scheme, as shown by the hPBE and hBLOC
points reported in Fig. \ref{fig5}
(compare also PBEint and hPBEint).

\section{Conclusions}
We performed a benchmark study of dihydrogen bond complexes.
Thus, we were able to define a set of reference geometries 
and interaction energies for a representative set of small 
complexes.
This set can be used in future assessments of methods for the
description of dihydrogen interactions.

In this work we have tested, against the benchmark, a few 
wave-function correlated methods. We found that second-order methods
(i.e. MP2 and QCISD) are rather accurate, giving mean absolute errors
of few tens of m\AA{} for H$\cdots$H bond lengths and 
about 0.4 kcal/mol for interaction energies. Nevertheless,
high accuracy appears to be out of reach for these approaches.
In particular, the MP2 method, although displaying a slightly better average
performance, generally shows a broad distribution
of the errors. Thus, it must be employed with caution because
relatively large errors can be obtained for some cases.
For this reason the use of the more reliable MP2.5
method may seem a good compromise between accuracy and
computational cost. Alternatively, we must acknowledge
the possibility of considering spin-resolved MP2 approaches 
(e.g. SCS- or SOS-MP2) \cite{scsmp2,scsmp2_rev,sosmp2}, 
eventually using a specialized parameterization \cite{scs_noncov,scspccp}, 
which already showed an encouraging performance for non-covalent 
interactions \cite{scsmp2_rev,scs_noncov,scspccp}.
Nevertheless, to this end a careful testing against
the benchmark must be considered in future work.

Finally, we had a survey on the performance of some popular
density functional methods, to understand the
level of accuracy that may be expected by such calculations.
Interestingly, we found that for the H$\cdots$H
bond length none of the functionals was able
to yield reliable results and a general overestimation
of the bond distance is found instead.
On the other hand, most functionals provide
quite accurate results for the interaction energies,
yielding a mean absolute error lower than 0.5 kcal/mol,
which is comparable to the MP2 and QCISD results.
However, the quality of the interaction energies
for single cases varies quite significantly, 
reflecting the broad differences between the
various dihydrogen complexes.
In fact, a detailed analysis of the density
and its related descriptors in the bonding region
of several complexes revealed that the different
features of the various dihydrogen bonds
can be hardly described at the semilocal level of the
theory. Thus, for a reliable description
of different complexes it appears necessary
to revert to higher rung functionals making
use of the occupied Kohn-Sham orbitals
(i.e. meta-GGAs and/or hybrids).

In conclusion, great caution shall be used when performing
DFT calculations on complexes displaying dihydrogen
bonding because DFT functionals appear generally unable
to fully describe the complex balancing of effects 
present in these systems. Nevertheless, meta-GGA
functionals and especially hybrids seem to give
higher reliability in this sense. Finally,
some attention must be payed to 
the possible mismatch between the
description of different properties and
functionals yielding a more balanced description
of different properties (see Fig. \ref{fig5}) shall be
 possibly preferred.

\section{Acknowledgments}
We thank TURBOMOLE GmbH for providing the TURBOMOLE program package.


\begin{thebibliography}{99}
%
\bibitem{sherrill13} Sherrill, C. D. \textit{Acc. Chem. Res.} \textbf{2013}, \textit{46}, 1020.
%
\bibitem{hohenstein12} Hohenstein, E. G.; Sherrill, C. D. \textit{WIREs Comput Mol Sci} \textbf{2012} \textit{2}, 304.
%
\bibitem{burns11} Burns, L. A.; V\'azquez-Mayagoitia, A.; Sumpter, B. G.; Sherrill, C. D. \textit{J. Chem. Phys.} \textbf{2011} \textit{134}, 084107.
%
\bibitem{thanthiriwatte11} Thanthiriwatte, K. S.; Hohenstein, E. G.; Burns, L. A.; Sherrill, C. D. \textit{ J. Chem. Theory Comput.} \textbf{2011} \textit{7}, 88.
%
\bibitem{sherrill09} Sherrill, C. D.; Takatani, T.; Hohenstein, E. G. \textit{J. Phys. Chem. A} \textbf{2009} \textit{113}, 10146.
%
\bibitem{dubecky13} Dubeck\'y, M.; Jure\v cka, P.; Derian, R.; Hobza, P.; Otyepka, M.; Mitas, L. \textit{J. Chem. Theory Comput.} \textbf{2013} \textit{9}, 4287.
%
\bibitem{sedlak} Sedlak, R.; Janowski, T.; Pito\v n\'ak, M.; \v Rez\'a\v c, J.; Pulay, P.; Hobza, P. \textit{J. Chem. Theory Comput.} \textbf{2013} \textit{9}, 3364.
%
\bibitem{hobza13} \v Rez\'a\v c, J.; Hobza, P. \textit{J. Chem. Theor. Comput.} \textbf{2013} \textit{9}, 2151.
%
\bibitem{melichercik13} Melicher\v c\'ik, M.; Pito\v n\'ak, M.; Kell\"o, V.; Hobza, P.; Neogr\'ady, P. \textit{ J. Chem. Theory Comput.} \textbf{2013} \textit{9}, 5296.
%
\bibitem{riley13} Riley, K. E.; Hobza, P. \textit{Phys. Chem. Chem. Phys.} \textbf{2013} \textit{15}, 17742.
%
\bibitem{riley13_2} Riley, K. E.; Murray, J. S.; Fanfrl\'ik, J.; \v Rez\'a\v c, J.; Sol\'a, R. J.; Concha, M. C.; Ramos, F. M.; Politzer, P. \textit{Journal of Molecular Modeling} \textbf{2013} \textit{19}, 4651. 
%
\bibitem{zhao07} Zhao, Y.; Truhlar, D. G. \textit{J. Chem. Theory Comput.} \textbf{2007}, \textit{3}, 289.
%
\bibitem{zhao06} Zhao, Y.; Truhlar, D. G. \textit{J. Chem. Theory Comput.} \textbf{2006} \textit{2}, 1009.
%
\bibitem{johnson13} Johnson, E. R.; Otero de la Roza, A.; Dale, S. G.; Di Labio, G. A. \textit{J. Chem. Phys.} \textbf{2013} \textit{139}, 214109.
%
\bibitem{otero13} Otero de la Roza, A.; Johnson, E. R. \textit{J. Chem. Phys.} \textbf{2013} \textit{138}, 204109.
%
\bibitem{johnson13_2} Johnson, E. R.; Salamone, M.; Bietti, M.; Di Labio, G. A. \textit{ J. Phys. Chem. A.} \textbf{2013} \textit{117}, 947.
%
\bibitem{contreras11} Contreras-Garc\'ia, J.; Johnson, E. R.; Keinan, S.; Chaudret, R.; Piquemal, J.-P.; Beratan,D. N.; Yang, W. \textit{J. Chem. Theory Comput.} \textbf{2011} \textit{7}, 625.
%
\bibitem{johnson10} Johnson, E. R.; Keinan, S.; Mori-S\'anchez, P.; Contreras-Garc\'ia, J.; Cohen, A. J.; Yang, W. \textit{J. Am. Chem. Soc.} \textbf{2010} \textit{132}, 6498.
%
\bibitem{grabowski13} Grabowski, S. J. \textit{Journal of Molecular Modeling} \textbf{2013} \textit{19}, 4713.
%
\bibitem{noncov_book} Hobza, P.; M\"uller-Dethlefs, K.; Jordan, K. D.; Lim, C. \textit{Non-Covalent Interactions: Theory and Experiment};  Royal Society of Chemistry: London, \textbf{2009}.
%
\bibitem{hbond_book1} Jeffrey, G. A. \textit{An Introduction to Hydrogen Bonding}; Oxford University Press: USA, \textbf{1997}.
%
\bibitem{hbond_book2} Grabowski, S. J. \textit{Hydrogen Bonding - New Insights}; Springer: Dordrecht, \textbf{2006}.
%
\bibitem{kollman} Kollman, P. A.; Allen, L. C. \textit{Chem. Rev.} \textbf{1972}, \textit{72}, 283.
%
\bibitem{zhao12} Zhao, G.-J.; Han, K. L. \textit{Acc. Chem. Res.} \textit{2012} \textbf{45}, 404.
%
\bibitem{grabowski11} Grabowski, S. J. \textit{Chem. Rev.} \textbf{2011} \textit{111}, 2597.
%
\bibitem{li11_2} Li, X.-Z.; Walker, B.; Michaelides, A. \textit{Proc. Natl. Accad. Soc.} \textbf{2011} \textit{108}, 6369.
%
\bibitem{contreras11_2} Contreras-Garc\'ia, J.; Yang, W.; Johnson, E. R. \textit{ J. Phys. Chem. A.} \textbf{2011} \textit{115}, 12983.
%
\bibitem{johnson09} Johnson, E. R.; Di Labio, G. A. \textit{Interdiscipl. Sci. - Comput. Life. Sci.} \textbf{2009} \textit{1}, 133.
%
\bibitem{grabowski13_2} Grabowski, S. J. \textit{Phys. Chem. Chem. Phys.} \textbf{2013} \textit{15}, 7249.
%
\bibitem{fuster11} Fuster, F.; Grabowski, S. J. \textit{J. Phys. Chem. A} \textbf{2011} \textit{115}, 10078.
%
\bibitem{dihydro_book} Bakhmutov, V. I. \textit{Dihydrogen Bonds: Principles, Experiments, and Applications}; John Wiley \& Sons, Inc.: Hoboken, New Jesrsey, \textbf{2008}.
%
\bibitem{dihydro_rev} Custelcean, R.; Jackson, J. E. \textit{Chem. Rev.} \textbf{2001} \textit{101}, 1963. 
%
\bibitem{grabowski04} Grabowski, S. J.; Sokalski, W. A.; Leszczynski, J. \textit{J. Phys. Chem. A} \textbf{2004} \textit{108}, 5823.
%
\bibitem{hu04} Hu, S.-W.; Wang, Y.; Wang, X.-Y.; Chu, T.-W.; Liu, X.-Q. \textit{J. Phys. Chem. A} \textbf{2004} \textit{108}, 1448.
%
\bibitem{hayashi05} Hayashi, A.; Shiga, M.; Tachikawa, M. \textit{Chem. Phys. Lett.} \textbf{2005} \textit{410}, 54.
%
\bibitem{solimannejad05} Solimannejad M.; Scheiner, S. \textit{J. Phys. Chem. A} \textbf{2005} \textit{109}, 6137.
%
\bibitem{alkorta06} Alkorta, I.; Zborowski, K.; Elguero, J.; Solimannejad, M. \textit{J. Phys. Chem. A} \textbf{2006} \textit{110}, 10279.
%
\bibitem{solimannejad06} Solimannejad M.; Alkorta, I. \textit{Chem. Phys.} \textbf{2006} \textit{324}, 459.
%
\bibitem{solimannejad06_2} Solimannejad M.; Boutalib, A. \textit{Chem. Phys.} \textbf{2006} \textit{320}, 275.
%
\bibitem{yao11} Yao A.; Ren, F. \textit{Comput. Theor. Chem.} \textbf{2011} \textit{963}, 463.
%
\bibitem{li11} Li, Y.; Zhang, L.; Du, S.; Ren, F.; Wang, W. \textit{Comput. Theor. Chem.} \textbf{2011} \textit{977}, 201.
%
\bibitem{meng05} Meng, Y.; Zhou, Z.; Duan, C.; Wang, B.; Zhong, Q. \textit{J. Mol. Struct. (Theochem)} \textbf{2005} \textit{713}, 135.
%
\bibitem{filippov06} Filippov, O. A.; Filin, A. M.; Tsupreva, V. N.; Belkova, N. V.; Lledos, A.; Uiaque, G.; Epstein, L. M.; Shubina, E. S. \textit{Inorg. Chem.} \textbf{2006} \textit{45}, 3086.
%
\bibitem{hugas07} Hugas, D.; Simon, S.; Duran, M. \textit{J. Phys. Chem. A} \textbf{2007} \textit{111}, 4506 (2007).
%
\bibitem{guo13} Guo, J.; Shi, V.; Ren, F.; Cao, D.; Zhang, Y. \textit{J. Mol. Model.} \textbf{2013} \textit{19}, 3153.
%
\bibitem{li13} Li, B.; Shi, W.; Ren, F. \textit{Comput. Theor. Chem.} \textbf{2013} \textit{1020}, 81.
%
\bibitem{grabowski13_3} Grabowski, S. J. \textit{J. Phys. Org. Chem.} \textbf{2013}, \textit{26}, 452.
%
\bibitem{filippov06_2} Filippov, O. A.; Filin, A. M.; Belkova, N. V.; Tsupreva, V. N.; Smirnova, Y. V.; Sivaev, I. B.; Epstein, L. M.; Shubina, E. S. \textit{J. Mol. Struct.} \textbf{2006} \textit{790}, 114. 
%
\bibitem{zhang11} Zhang, H.; Li, X.; Tang, Y. \textit{Front. Phys.} \textbf{2011} \textit{6}, 213.
%
\bibitem{sandhya12} Sandhya K. S.; Suresh, C. H. \textit{Dalton Trans.} \textbf{2012} \textit{41}, 11018.
%
\bibitem{flener12} Flener Lovitt, C.; Frenking, G.; Girolami, G. S. \textit{Organometallics} \textbf{2012} \textit{31}, 4122.
%
\bibitem{yang12} Yang, X. \textit{J. Clust. Sci.} \textbf{2012} \textit{23}, 703.
%
\bibitem{grabowski00} Grabowski, S. J. \textit{J. Phys. Chem. A} \textbf{2000} \textit{104}, 5551.
%
\bibitem{ccsdt} Raghavachari, K.; Trucks, G. W.; Pople, J. A.; Head-Gordon, M. \textit{Chem. Phys. Lett.} \textbf{1989} \textit{157}, 479.
%
\bibitem{qcisd} Pople, J. A.; Head-Gordon, M.; Raghavachari, K. \textit{J. Chem. Phys.} \textbf{1987} \textit{87}, 5968.
%
\bibitem{qcisdt_1} Gauss J.; Cremer, D. \textit{Chem. Phys. Lett.} \textbf{1988} \textit{150}, 280.
%
\bibitem{qcisdt_2} Salter, E. A.; Trucks, G. W.; Bartlett, R. J. \textit{J. Chem. Phys.} \textbf{1989} \textit{90}, 1752.
%
\bibitem{mp} M\o ller C.; Plesset, M. S. \textit{Phys. Rev.} \textbf{1934} \textit{46}, 0618.
%
\bibitem{mp2} Head-Gordon, M.; Pople, J. A.; Frisch, M. J. \textit{Chem. Phys. Lett.} \textbf{1988} \textit{153}, 503.
%
\bibitem{mp2.5} Pito\v n\'ak, M.; Neogr\'ady, P.; \v Cern\'y, J.; Grimme, S.; Hobza, P. \textit{ChemPhysChem} \textbf{2009} \textit{10}, 282.
%
\bibitem{mp4} Raghavachari K.; Pople, J. A. \textit{Int. J. Quantum Chem.} \textbf{1978} \textit{14}, 91.
%
\bibitem{hohenstein10} Hohenstein, E. G.; Sherrill, C. D. \textit{J. Chem. Phys.} \textbf{2010} \textit{133}, 014101.
%
\bibitem{slater51} Slater, J. C. \textit{Phys. Rev.} \textbf{1951} \textit{81}, 385.
%
\bibitem{dirac29} Dirac, P. A. M. \textit{Proc. Royal Soc. (London) A} \textbf{1929} \textit{123}, 714.
%
\bibitem{vwn} Vosko, S. H.; Wilk, L.; Nusair, M. \textit{Can. J. Phys.} \textbf{1980} \textit{58}, 1200.
%
\bibitem{b88} Becke, A. D. \textit{Phys. Rev. A} \textbf{1988} \textit{38}, 3098.
%
\bibitem{p86} Perdew, J. P. \textit{Phys. Rev. B} \textbf{1986} \textit{33}, 8822.
%
\bibitem{lyp} Lee, C.; Yang, W.; Parr, R. G. \textit{Phys. Rev. B} \textbf{1988} \textit{37}, 785.
%
\bibitem{optx} Handy N. C.; Cohen, A. J. \textit{Mol. Phys.} \textbf{2001} \textit{99}, 403.
%
\bibitem{pbe} Perdew, J. P.; Burke, K.; Ernzerhof, M. \textit{Phys. Rev. Lett.} \textbf{1996} \textit{77}, 3865.
%
\bibitem{pbeint} Fabiano, E.; Constantin, L. A.; Della Sala, F. \textit{Phys. Rev. B} \textbf{2010} \textit{82}, 113104.
%
\bibitem{subroutines} FORTRAN90 routines are freely available at http://www.theory-nnl.it/software.php; accessed on April 2014.
%
\bibitem{apbe} Constantin, L. A.; Fabiano, E.; Laricchia, S.; Della Sala, F. \textit{Phys. Rev. Lett.} \textbf{2011} \textit{106}, 186406.  
%
\bibitem{tpss} Tao, J.; Perdew, J. P.; Staroverov, V. N.; Scuseria, G. E. \textit{Phys. Rev. Lett.} \textbf{2003} \textit{91}, 146401.
%
\bibitem{revtpss} Perdew, J. P.; Ruzsinszky, A.; Csonka, G. I.; Constantin, L. A.; Sun, J. \textit{Phys. Rev. Lett.} \textbf{2009} \textit{103}, 026403; \textit{Phys. Rev. Lett.} \textbf{2011} \textit{106}, 179902.
%
\bibitem{bloc} Constantin, L. A.; Fabiano, E.; Della Sala, F. \textit{J. Chem. Theory Comput.} \textbf{2013} \textit{9}, 2256.
%
\bibitem{loc} Constantin, L. A.; Fabiano, E.; Della Sala, F. \textit{Phys. Rev. B} \textbf{2012} \textit{86}, 035130. 
%
\bibitem{bhole} Constantin, L. A.; Fabiano, E.; Della Sala, F. \textit{Phys. Rev. B} \textbf{2013} \textit{88}, 125112.
%
\bibitem{vsxc} Van Voorhis, T.; Scuseria, G. E. \textit{J. Chem. Phys.} \textbf{1998} \textit{109}, 400.
%
\bibitem{m06l} Zhao Y.; Truhlar, D. G. \textit{J. Chem. Phys.} \textbf{2006} \textit{125}, 194101.
%
\bibitem{b3lyp_1} Becke, A. D. \textit{J. Chem. Phys.} \textbf{1993} \textit{98}, 5648.
%
\bibitem{b3lyp_2} Stephens, P. J.; Devlin, F. J.; Chabalowski, C. F.; Frisch, M. J. \textit{J. Phys. Chem.} \textbf{1994} \textit{98}, 11623.
%
\bibitem{bhlyp} Becke, A. D. \textit{J. Chem. Phys.} \textbf{1993} \textit{98}, 1372.
%
\bibitem{o3lyp} Cohen A. J.; Handy, N. C. \textit{Mol. Phys.} \textbf{2001} \textit{99}, 607.
%
\bibitem{pbe0} Adamo, C.; Barone, V. \textit{J. Chem. Phys.} \textbf{1999} \textit{110}, 6158.
%
\bibitem{hpbeint} Fabiano, E.; Constantin, L. A.; Della Sala, F. \textit{Int. J. Quantum Chem.} \textbf{2013} \textit{113}, 673.
%
\bibitem{tpssh} Staroverov, V. N.; Scuseria, G. E.; Tao, J.; Perdew, J. P. \textit{J. Chem. Phys.} \textbf{2003} \textit{119}, 12129.
%
\bibitem{m06} Zhao Y.; Truhlar, D. G. \textit{Theor. Chem. Acc.} \textbf{2008} \textit{120}, 215.
%
\bibitem{m06hf} Zhao Y.; Truhlar, D. G. \textit{J. Phys. Chem. A} \textbf{2006} \textit{110}, 13126.
%
\bibitem{yanai04} Yanai, T.; Tew, D. P.; Handy. N. C. \textit{Chem. Phys. Lett.} \textbf{2004} \textit{393}, 51.
%
\bibitem{tawada04} Tawada, Y.; Tsuneda, T.; Yanagisawa, S.; Yanai, T.; Hirao. K. \textit{J. Chem. Phys.} \textbf{2004} \textit{120}, 8425.
%
\bibitem{chai08} Chai, J.-D.; Head-Gordon. M. \textit{J. Chem. Phys.} \textbf{2008} \textit{128}, 084106.
%
\bibitem{aug-cc-pvtz1} Dunning Jr., T. H. \textit{J. Chem. Phys.} \textbf{1989} \textit{90}, 1007.
%
\bibitem{aug-cc-pvtz2} Woon D. E.; Dunning Jr., T. H. \textit{J. Chem. Phys.} \textbf{1993} \textit{98}, 1358.
%
\bibitem{aug-cc-pvtz3} Kendall, R. A.; Dunning Jr., T. H.; Harrison, R. J. \textit{J. Chem. Phys.} \textbf{1992} \textit{96}, 6796.
%
%
\bibitem{danovich13} Danovich, D.; Shaik, S.; Neese, F.; Echeverr\'ia, J.; Aull\'on, G.; Alvarez S. \textit{J. Chem. Theory Comput.} \textbf{2013} \textit{9}, 1977.
%
\bibitem{halkier98} Halkier, A.; Helgaker, T.; J\o rgensen, P.; Klopper, W.; Koch, H.; Olsen, J.; Wilson, A. K. \textit{Chem. Phys. Lett.} \textbf{1998} \textit{286}, 243.
%
\bibitem{myrpa} Fabiano, E.; Della Sala, F. \textit{Theor. Chem. Acc.} \textbf{2012} \textit{131}, 1278. 
%
\bibitem{east93} East, A. L. L.; Allen, W. D. \textit{J. Chem. Phys.} \textbf{1993} \textit{99}, 4638.
%
\bibitem{csaszar98} Csaszar, A. G.; Allen, W. D.; Schaefer, H. F. \textit{J. Chem. Phys.} \textbf{1998} \textit{108}, 9571.
%
\bibitem{burns14} Burns, L. A.; Marshall, M. S.; Sherrill, C. D. \textit{J. Chem. Theory Comput.} \textbf{2014} \textit{10}, 49.
%
\bibitem{mackie11} Mackie, I. D.; DiLabio, G. A. \textit{J. Chem. Phys.} \textbf{2011} \textit{135}, 134318.
%
\bibitem{feller11} Feller, D.; Peterson, K. A.; Grant Hill, J. \textit{J. Chem. Phys.} \textbf{2011} \textit{135}, 044102.
%
\bibitem{def2tzvpp1} Weigend, F.; Furche, F.; Ahlrichs, R. \textit{J. Chem. Phys.} \textbf{2003} \textit{119}, 12753.
%
\bibitem{def2tzvpp2} Weigend, F.; Ahlrichs, R. \textit{Phys. Chem. Chem. Phys.} \textbf{2005} \textit{7}, 3297.
%
\bibitem{cp} Boys S. F.; Bernardi, F. \textit{Mol. Phys.} \textbf{1970} \textit{19}, 553.
%
\bibitem{turbomole} TURBOMOLE, V6.3; TURBOMOLE GmbH: Karlsruhe, Germany, 2011. Available from http://www.turbomole.com (accessed April 2014).
%
\bibitem{orca} Neese, F. \textit{WIREs Comput. Mol. Sci.} \textbf{2012} \textit{2}, 73.
%
\bibitem{psi4} Turney, J. M.; Simmonett, A. C.; Parrish, R. M.; Hohenstein, E. G.; Evangelista, F.; Fermann, J. T.; Mintz, B. J.; Burns, L. A.; Wilke, J. J.; Abrams, M. L.; Russ, N. J.; Leininger, M. L.; Janssen, C. L.; Seidl, E. T.; Allen, W. D.; Schaefer, H. F.; King, R. A.; Valeev, E. F.; Sherrill, C. D.;Crawford, T. D. \textit{WIREs Comput. Mol. Sci.} \textbf{2012} \textit{2}, 556.
%
%
\bibitem{feller06} Feller D.; Peterson, K. A. \textit{J. Chem. Phys.} \textbf{2006} \textit{124}, 054107.
%
\bibitem{singh09} Singh, N. J.; Min, S. K.; Kim, D. Y.; Kim, K. S. \textit{J. Chem. Theory Comput.} \textbf{2009} \textit{5}, 515.
%
\bibitem{hoja14} Hoja, J.; Sax, A. F.; Szalewicz, K. \textit{Chem. Eur. J.} \textbf{2014} \textit{20}, 2292.
%
\bibitem{arey09} Arey, J. S.; Aeberhard, P. C.; Lin, I.-C.; Rothlisberger, U. \textit{J. Phys. Chem. B} \textbf{2009} \textit{113}, 4726.
%
%
\bibitem{hujo11} Hujo, W.; Grimme, S. \textit{Phys. Chem. Chem. Phys.} \textbf{2011} \textit{13}, 13942.
%
\bibitem{dilabio13} Di Labio, G. A.; Johnson, E. R.; Otero de la Roza, A. \textit{Phys. Chem. Chem. Phys.} \textbf{2013} \textit{15}, 12821.

\bibitem{hb06} Zhao Y.; Truhlar, D. G. \textit{J. Chem. Theory Comput.} \textbf{2005} \textit{1}, 415.
%
\bibitem{nci} Contreras-Garc\'ia, J.; Johnson, E. R.; Keinan, S.; Chaudret, R.; Piquemal, J.-P.; Beratan, D. N.; Yang, W. \textit{J. Chem. Theory Comput.} \textbf{2011} \textit{7}, 625.
%
\bibitem{nci2} Contreras-Garc\'ia, J.; Yang, W.; Johnson, E. R. \textit{J. Phys. Chem. A} \textbf{2011} \textit{115}, 12983.
%
\bibitem{fdeene} Laricchia, S. ; Fabiano, E.; Della Sala, F. \textit{J. Chem. Phys.} \textbf{2012} \textit{137}, 014102. 
%
\bibitem{fdect} Laricchia, S.; Fabiano, E.; Della Sala, F. \textit{J. Chem. Phys.} \textbf{2013} \textit{138}, 124112. 
%
\bibitem{mukappa} Fabiano, E.; Constantin, L. A.; Della Sala, F. \textit{J. Chem. Theory Comput.} \textbf{2011} \textit{7} , 3548
%
\bibitem{scsmp2} Grimme, S. \textit{J. Chem. Phys.} \textbf{2003} \textit{118}, 9095.
%
\bibitem{scsmp2_rev} Grimme, S.; Goerigk L.; Fink, R. F. \textit{Wiley Interdiscip. Rev.: Comput. Mol. Sci.} \textbf{2012} \textit{2}, 886.
%
\bibitem{sosmp2} Jung, Y.; Lochan, R. C.; Dutoi A. D.; Head-Gordon, M. \textit{J. Chem. Phys.} \textbf{2004} \textit{121}, 9793.
%
\bibitem{scs_noncov} Distasio Jr., R. A.; Head-Gordon, M. \textit{Mol. Phys.} \textbf{2007} \textit{105}, 1073.
%
\bibitem{scspccp} Grabowski, I.; Fabiano, E.; Della Sala, F. \textit{Phys. Chem. Chem. Phys.} \textbf{2013} \textit{15}, 15485. 
%
\end{thebibliography}
\end{document}